\documentclass{mem}
\usepackage{natbib}\usepackage{txfonts}\usepackage{balance}
\usepackage{graphicx}
\usepackage{epstopdf}
\pdfoutput=1
\usepackage[a4paper,breaklinks]{hyperref}
\idline{0}{0}
\begin{document}
\def\teff{$T\rm_{eff }$}
\def\kms{$\mathrm {km s}^{-1}$}

\title{Constraints on mass loss of globular clusters in dwarf galaxies}
\subtitle{}

\author{
S.\ S.\ Larsen  \inst{1} 
\and
J.\ Strader \inst{2}
\and 
J.\ P.\ Brodie \inst{3}
}

\offprints{S.\ S.\ Larsen}

\institute{
Department of Astrophysics / IMAPP,
Radboud University Nijmegen, 
P.O.\ Box 9010, 6500 GL Nijmegen, The Netherlands \\
\email{s.larsen@astro.ru.nl}
\and
Department of Physics and Astronomy, Michigan State University, East Lansing, Michigan 48824, USA
\and
UCO/Lick Observatory, University of California, Santa Cruz, CA 95064, USA
}

\authorrunning{Larsen }

\titlerunning{Globular Clusters in Dwarf Galaxies}

\abstract{
The Fornax dwarf spheroidal galaxy is well known for its very high globular cluster specific frequency, $S_N\approx26$. 
Furthermore, while the field star metallicity distribution peaks at $\mathrm{[Fe/H]}\approx-1$, four of the five GCs have $\mathrm{[Fe/H]}<-2$. Only about 5\% of the field stars have such low metallicities. Hence, a very large fraction of about 1/5-1/4 of the most metal-poor stars belong to the four most metal-poor GCs. This implies that these clusters could, at most, have been a factor of $\sim4-5$ more massive initially. A second, even more extreme case may be the IKN dwarf galaxy where $S_N\approx124$. Although  metallicities are not accurately known, the GCs account for about 13\% of the \emph{total} $V$-band luminosity of IKN.
}
\maketitle{}

\section{Introduction}

One of the major puzzles related to the phenomenon of multiple stellar populations in globular clusters (GCs) is the ``mass budget problem''. Half or more of the present-day stellar mass in GCs typically resides in the 2nd generation, whose anomalous chemical abundances suggest that it formed out of ejecta produced by the first generation of stars. In most scenarios, this requires that the first generation was initially much more populous than it is today, by perhaps an order of magnitude or more \citep[e.g.][]{Gratton2012}. 
A test of whether such scenarios are viable would therefore be to determine how many stars in a galaxy can be traced back to GCs. In the Milky Way, this may be quite difficult in practice \citep[but perhaps  worth contemplating e.g.\ via chemical tagging;][]{Hawthorn2010}. However,  dwarf galaxies may hold promising potential for carrying out this test. Although their GC populations are relatively modest in absolute terms, the few GCs that are present in some dwarfs can account for a substantial fraction of the total number of stars in the galaxy. Even without detailed constraints on the chemical composition of the stars within and out of the GCs, the ratio of GCs to field stars in a suitably restricted metallicity interval may thus be sufficiently high to provide useful constraints on how much mass the clusters could have lost.

The GC specific frequency, $S_N$\footnote{$S_N = N_\mathrm{GC} \times 10^{0.4 (M_V - 15)}$ where $N_\mathrm{GC}$ is the number of GCs in a galaxy and $M_V$ the absolute $V$ magnitude of the galaxy \citep{Harris1981}}, varies substantially among galaxies, with spiral galaxies typically having $S_N\approx 1$ and elliptical galaxies having $S_N\approx$ 3--5. Even richer GC systems are found around cD galaxies, which can have $S_N$ up to 10--15. 
The highest GC specific frequencies are found in dwarf galaxies, of which the Fornax dSph is a  well-known example. For $M_V=-13.2$ \citep{Mateo1998}, its five GCs \citep{Hodge1961} correspond to $S_N = 26$. Fornax does not appear to be an outlier or even particularly extreme in this respect; other examples of dwarf galaxies with very high specific frequencies are listed in \citet{Peng2008} and \citet{Georgiev2010}.

These global numbers hide the fact that individual stellar sub-populations within galaxies can have much higher GC specific frequencies. In the Milky Way, about 2/3 of the 150 ancient GCs currently known \citep{Harris1996} are associated kinematically, chemically and spatially with the halo, although this component only accounts for about 1\% of the stellar mass in our Galaxy. This ``second specific frequency'' problem is not restricted to our Galaxy; also in elliptical galaxies a disproportionately large fraction of the GCs appear to be associated with the metal-poor stars \citep{Forte1981,Harris2002}. However,  in the Milky Way halo, the GCs still only \emph{currently} account for $\sim2$\% of the stellar mass \citep{Kruijssen2008}. This fraction may, however, have been very much higher in the past.

\section{Abundances of  stars and GCs in Fornax}

The Fornax dSph exhibits the second specific frequency problem. The field star metallicity distribution has a broad peak around $\mathrm{[Fe/H]}\approx-1$ \citep{Battaglia2006,Kirby2011} with only a small fraction of the stars having $\mathrm{[Fe/H]}<-2$. In contrast,  four of the five GCs have low metallicities around $\mathrm{[Fe/H]}\approx-2$. The specific frequency of the GCs, relative to field stars of the same metallicity, must therefore be very high indeed.

Accurate metallicity determinations  are now available both for field stars and all the GCs in the Fornax dSph,  allowing a more detailed analysis. \citet{Battaglia2006} published measurements of the Ca {\sc ii} IR triplet for 562 field red giants, covering the full radial range ($\sim1^\circ$). A comparison with [Fe/H] abundances derived from high-dispersion spectroscopy for a subset of the stars showed that the Ca {\sc ii} triplet measurements are reliable proxies for the Fe abundances (within 0.1--0.2 dex) for $\mathrm{[Fe/H]}>-2.5$ \citep{Battaglia2007}. Measurements of  [Fe/H] abundances by direct spectral fitting \citep{Kirby2011} yield a very similar metallicity distribution to that derived by \citet{Battaglia2006}.

\begin{table}
%\begin{minipage}[t]{85mm}
\caption{Iron abundances and $[$Ca/Fe$]$ ratios  for the five GCs in the Fornax dSph from high-dispersion spectroscopy.
}
\label{tab:fornaxgc}
\begin{center}
\begin{tabular}{lccl} \hline
                 & [Fe/H]                &   [Ca/Fe]  & Source \\ \hline
Fornax 1 & $-2.5\pm0.1$    & $+0.15\pm0.04$ & L06 \\
Fornax 2 & $-2.1\pm0.1$    & $+0.20\pm0.03$ &  L06 \\
Fornax 3 & $-2.3\pm0.1$    & $+0.25\pm0.08$ & L12 \\
Fornax 4 & $-1.4\pm0.1$    & $+0.13\pm0.07$ & L12 \\
Fornax 5 & $-2.1\pm0.1$    & $+0.27\pm0.09$ & L12 \\
\hline
\end{tabular}
\end{center}
\end{table}

Abundance measurements from high-dispersion spectroscopy of individual stars are available for three of the GCs  (Fornax 1, 2 and 3) \citep[][L06]{Letarte2006}. 
We have recently carried out a detailed abundance analysis based on high-dispersion spectroscopy of \emph{integrated} light for Fornax 3, 4 and 5 \citep[][L12]{Larsen2012a}. For Fornax 3 our [Fe/H] abundance agreed with that measured by Letarte et al.\ within 0.1 dex. For Fornax 5 we found a low metallicity ([Fe/H]=$-2.1$) while Fornax 4 was confirmed to be more metal-rich than the other four clusters ([Fe/H]=$-1.4\pm0.1$). The high-dispersion spectroscopic Fe and Ca abundances are summarized in Table~\ref{tab:fornaxgc}. All five GCs have somewhat super-solar [Ca/Fe] abundance ratios, in agreement with measurements of field star abundances for  Fornax  and other dSphs \citep{Tolstoy2009}.

\section{Stars in the field vs.\ GCs}

\begin{figure}[]
\resizebox{\hsize}{!}{\includegraphics[clip=true]{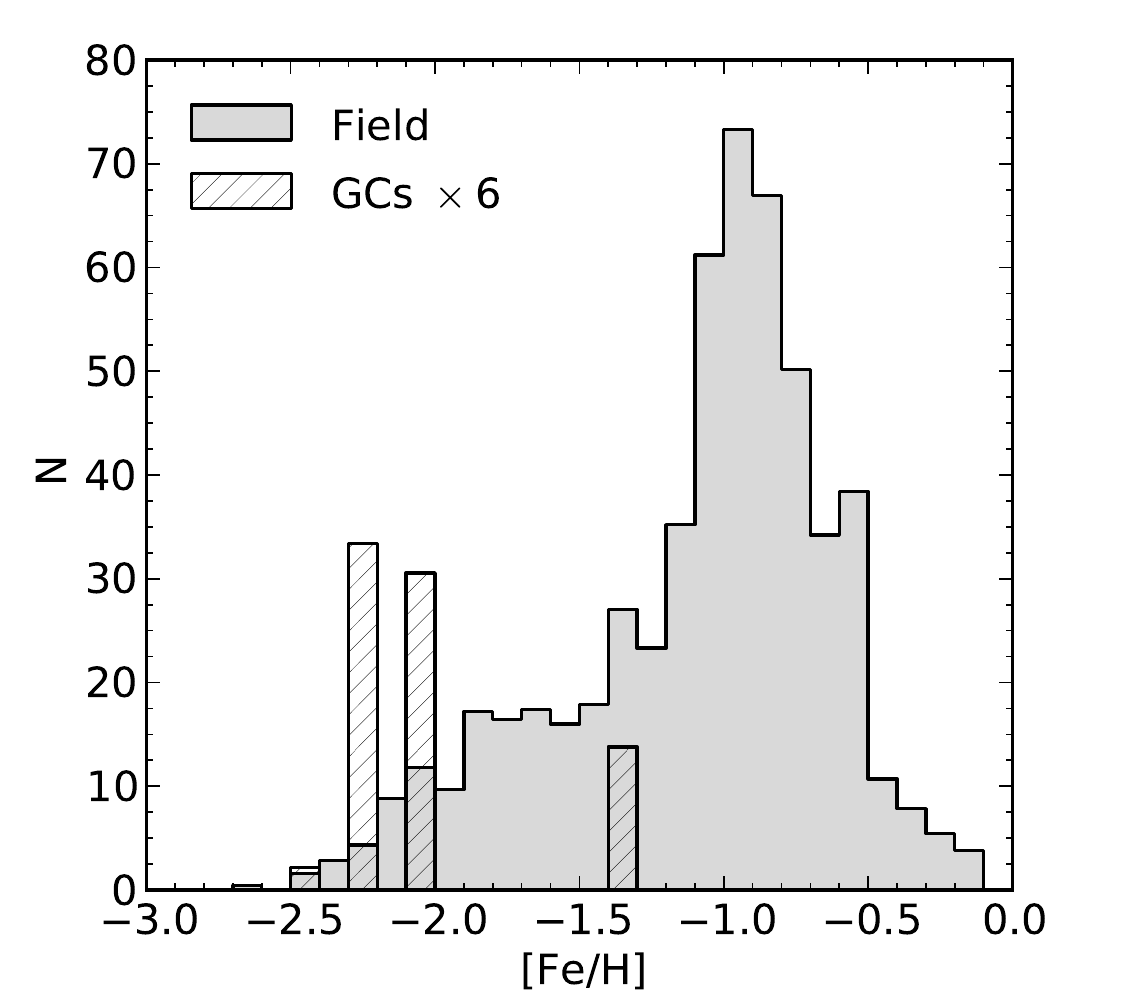}}
\caption{
\footnotesize
Distribution of [Fe/H] abundances for field stars and GCs in the Fornax dSph.}
\label{fig:zglob}
\end{figure}

We corrected the Battaglia et al.\ metallicity distribution for spatial coverage and also took into account that the relative number of RGB stars above the spectroscopic magnitude limit  depends on age and metallicity. Details are in \citet{Larsen2012}. 
Figure~\ref{fig:zglob} shows the global metallicity distribution of the field stars, corrected for these selection effects. Also shown is the metallicity distribution of the GCs, where each GC has been counted as $6\times N_\mathrm{stars} \times 10^{-0.4 (M_{V,\mathrm{GC}} + 13.2)}$. Here, $M_V=-13.2$ is the  $M_V$ magnitude of the Fornax dSph, $M_{V,\mathrm{GC}}$ are the $M_V$ magnitudes of the individual GCs, and $N_\mathrm{stars} = 562$ is the number of stars in the Battaglia et al.\ sample. Thus, the scale of the GCs is exaggerated by about a factor of 6.

It is immediately obvious from Fig.~\ref{fig:zglob} that the metallicity distribution of the GCs differs enormously from that of the field stars. 

As a first  estimate of the fraction of metal-poor stars associated with the GCs, we simply scale the total luminosity of the Fornax dSph by the fraction of stars with $\mathrm{[Fe/H]}<-2$. With the corrections mentioned above, this fraction is 5\%, so that the integrated absolute magnitude of these stars is $M_{V} = -10.0$. The integrated magnitude of the four metal-poor GCs is $M_{V,\mathrm{GCs}} = -8.9$ \citep{Webbink1985}. In other words, \emph{the four metal-poor GCs account for more than 1/4 of the luminosity of all stars with metallicities $\mathrm{[Fe/H]}<-2$}. A correction for the age- and metallicity dependent mass-to-light ratio of the field stars tends to further increase this fraction by 10\%--20\% \citep{Larsen2012}. This comparison is largely independent of assumptions about the IMF.

We can also estimate the fraction of stars by \emph{mass} that reside in the metal-poor GCs, although this involves more assumptions. Recent estimates of the  stellar mass of the Fornax dSph range from $4.3\times10^7 M_\odot$ to $6.1\times10^7 M_\odot$ \citep{deBoer2012,Coleman2008}. The mean virial $M/L_V$ ratio of the metal-poor GCs is estimated to be $M/L_V\approx3.5$, corresponding to a total mass of $1.0\times10^6 M_\odot$ for these GCs \citep{Larsen2012a}. After correcting for age- and metallicity effects, and conservatively adopting the larger of the above estimates for the  stellar mass of Fornax, we again find that a fraction of 1/5-1/4 of the metal-poor stars in Fornax are associated with the four metal-poor GCs. This comparison is, however, complicated by the relatively high observed $M/L_V$ ratio of the GCs; a lower value of $M/L_V\approx2$ is predicted for a standard Kroupa- or Chabrier-type IMF at these metallicities. If we were to adopt this lower $M/L_V$ ratio for the GCs (albeit at odds with the dynamical measurements), the mass fraction of metal-poor stars in the GCs would decrease to 1/9-1/7.

\section{Dwarf Galaxies and Multiple Populations in GCs}

Regardless of the uncertainties in the above analysis, it is clear that a very large fraction of the metal-poor stars in the Fornax dSph are \emph{currently} -- even after a Hubble time -- members of GCs. 
The implication is that the Fornax GCs could not have been much more than a factor of 4--5 more massive initially -- there simply aren't enough metal-poor stars present in the galaxy today to account for any more mass loss from the GCs. Furthermore, this must be considered a conservative upper limit as it assumes that no other stars with $\mathrm{[Fe/H]}<-2$ formed in the field, or in other clusters that have since dissolved.

One potential caveat is that stars may have been lost from the Fornax dSph over its lifetime. However, \citet{Penarrubia2009} studied the spatial profiles of several dSph galaxies in the Local Group, and concluded, based on N-body simulations of the signatures of tidal stripping that ``The Fornax dwarf provides probably the clearest case of a dwarf whose stellar component has \emph{not} been disturbed by Galactic tides'' (their emphasis). 

The issue would become moot if the Fornax GCs do not host the multiple populations that appear to be so ubiquitous in Galactic GCs. This, by itself, would of course be interesting -- it would imply that GC formation in dwarf galaxies is quite a different process from that in larger galaxies (or, at least, the Milky Way). It is therefore worthwhile examining the constraints on the presence of multiple stellar populations in the Fornax GCs. The challenge here is that the clusters are too far away to easily observe large numbers of individual member stars spectroscopically. However, in their observations of 3 RGB stars in each of the clusters Fornax 1, 2 and 3, \citet{Letarte2006} found at least one star (in Fornax 3) which appears to have depleted Mg and O and enhanced Na,  suggesting that the chemical abundance anomalies are present at least in this cluster. In our analysis of the integrated-light spectra, we found Mg to depleted relative to Ca and Ti in Fornax 3, 4 and 5, possibly a hint of the Mg-Al anticorrelation. While these data suggest that the Fornax GCs are not very different from their Milky Way counterparts, stronger constraints on the presence (or absence) of multiple stellar populations are clearly desirable. 

Another question is whether other cases like Fornax can be found. One such case may be the IKN dwarf spheroidal in the Ursa Major group. This galaxy is even fainter than Fornax \citep[$M_V = -11.5$;][]{Georgiev2010} but also hosts 5 GCs. This makes the GC specific frequency of IKN even higher than that of Fornax, at a staggering $S_N = 124$. Photometry of RGB stars in IKN reveals a broad metallicity distribution reminiscent of that in Fornax \citep{Lianou2010}. The metallicities of the GCs are not accurately known, but their blue integrated colours suggest relatively low metallicities \citep{Georgiev2010}. However, without even trying to match the metallicities of GCs and field stars, the GCs account for about 13\% of the \emph{total} integrated $V$-band luminosity of the IKN galaxy.
 
\begin{acknowledgements}
SSL is  grateful to the Leids Kerkhoven-Bosscha Fonds for supporting his attendance of this conference by means of a travel grant.
\end{acknowledgements}

\bibliographystyle{aa}

\end{document}